\documentclass[aps,prl,superscriptaddress,showpacs,twocolumn]{revtex4}
\usepackage{epsfig}
\begin{document}
\title{Experimental evidence for phase synchronization transitions in human 
cardio-respiratory system}
\author{Ronny Bartsch}
\affiliation{Minerva Center, Department of Physics, Bar-Ilan University,
  Ramat-Gan 52900, Israel}
\author{Jan W. Kantelhardt}
\affiliation{Intitute of Physics, Theory Group,
  Martin-Luther-Universit\"at Halle-Wittenberg, 06099 Halle (Saale), Germany }
\author{Thomas Penzel}
\affiliation{Division of Pulmonary Diseases, Department of Internal Medicine,
  Hospital of Philipps-University, 35033 Marburg, Germany}
\author{Shlomo Havlin}
\affiliation{Minerva Center, Department of Physics, Bar-Ilan University,
  Ramat-Gan 52900, Israel}
\date{submitted April 3, 2006}

\begin{abstract} 
Transitions in the dynamics of complex systems can be characterized by changes 
in the synchronization behavior of their components.  Taking the human cardio-respiratory 
system as an example and using an automated procedure for screening the synchrograms 
of 112 healthy subjects we study the frequency and the distribution of synchronization 
episodes under different physiological conditions that occur during sleep.  
We find that phase synchronization between heartbeat and breathing is significantly 
enhanced during non-rapid-eye-movement (non-REM) sleep (deep sleep and light sleep) 
and reduced during REM sleep.  
Our results suggest that the synchronization is mainly due to a weak influence of the 
breathing oscillator upon the heartbeat oscillator, which is disturbed in the presence 
of long-term correlated noise, superimposed by the activity of higher brain regions 
during REM sleep.
\end{abstract}
 \maketitle

Periodic events are ubiquitous in many natural systems \cite{Glass}.  If 
two oscillatory processes are weakly coupled, they can become phase synchronized.  
Transitions in the synchronization behavior have been shown to be important 
characteristics of coupled oscillatory model systems \cite{Zusatz}.  It has also 
been found that noise, when applied identically to different nonlinear oscillators, 
can induce, enhance, or destroy synchronization among them \cite{noise,chem,laser}.
However, phase synchronization is difficult to study in experimental data which are 
very often inherently nonstationary and thus contain only quasiperiodic oscillations.  
Among the few recent experimental studies are coupled electrochemical oscillators 
\cite{chem}, laser systems \cite{laser}, and climate variables \cite{geophys}.  
In physiology, the study of phase synchronization focusses on cardio-respiratory 
data (see below) and encephalographic data \cite{encephalography}.  Here, in order 
to obtain reliable experimental evidences of transitions in phase synchronization 
behavior, we consider cardio-respiratory synchronization in humans during sleep, 
because homogeneous long-term data for well defined conditions of a complex system 
is available in this particular example.

First approaches for the study of cardio-respiratory synchronization have been
undertaken by the analysis of the relative position of inspiration within the 
corresponding cardiac cycle \cite{Engel}.  More recently, phase synchronization 
between heartbeat and breathing has been studied during wakefulness using the 
synchrogram method \cite{tutorial,Schaefer,Toledo,Stefanovska}.  While long 
synchronization episodes were observed in athletes and heart transplant patients 
(several hundreds of seconds) \cite{Schaefer,Toledo}, shorter episodes were
detected in normal subjects (typical duration less than hundred seconds)
\cite{Toledo,Stefanovska,Prokhorov}.  For two recent models of cardio-respiratory 
synchronization, see \cite{Kotani,Smely}.

In this Letter we use the concept of phase synchronization to develop an automated
synchrogram based procedure and study interactions between cardiac and respiratory
oscillations under different well-defined physiological conditions.  We focus on the 
sleep stages, where external stimuli are absent.  It is well known that healthy sleep 
consists of cycles of roughly 1-2 hours duration.  Each cycle is characterized by a 
sequence starting usually with light sleep, followed by deep sleep and REM sleep 
(rapid eye movement) \cite{rechtschaffen}.  We find the intriguing result that 
during REM sleep cardio-respiratory synchronization is suppressed by approximately a 
factor of 3 compared with wakefulness.  On the other hand, during non-REM sleep, it 
is enhanced by a factor of 2.4, again compared with wakefulness.  In addition,
we find that these significant differences between synchronization in REM and
non-REM sleep are very stable and occur in the same way for males and females,
independent of age and independent of the body mass index (BMI).  Hence it seems
likely that -- similar to the long-term correlations \cite{Boston} occurring in 
both heartbeat \cite{HBcor} and breathing \cite{BRcor} during REM sleep but not 
during non-REM sleep -- the differences are caused by the influence of the activity 
of higher brain regions on both oscillators \cite{fn1}.

First, we developed an algorithm, which detects epochs of synchronization 
automatically and systematically.  The algorithm is applied on simultaneous records 
of respiration (from a thermistor placed close to the subject's nose) and heartbeat
(from electrocardiograms) obtained for 112 healthy subjects during sleep.  
The data was recorded 
in the EU project SIESTA in several European sleep laboratories \cite{Siesta}, and 
the average length of the records is 7.9 hours with a standard deviation of 25 
minutes. Sleep stages have been determined by visual evaluation of electrophysiological 
recordings of brain activity \cite{rechtschaffen}.  We can thus assign the 
synchronization episodes to the specific sleep stages.  In addition, we constructed 
surrogate data by random combination of heartbeat and breathing signals from different 
subjects \cite{Toledo}.  We found that the total duration of the detected 
synchronization episodes in real data is increased by a factor 2.3 as compared with 
the surrogate data, suggesting that most of the detected episodes are real 
\cite{spurious}.

\begin{figure} \begin{center} \epsfig{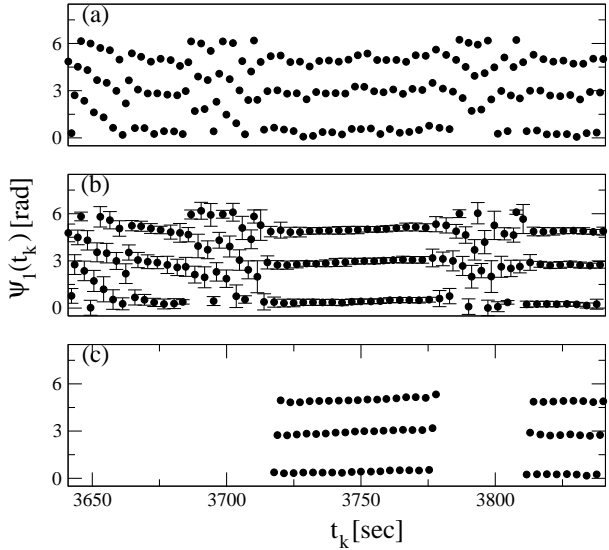}
\end{center} \caption{Illustration of our automated phase synchronization detector:
(a) section of a typical synchrogram in raw form, (b) the same section as in (a)
after applying a moving average filter with windows of $\tau=30$ s around each
breathing cycle; the window averages are shown together with their standard deviations
(error bars), (c) contiguous phase points of (b) where the standard deviation is
below a threshold (see text) are detected as synchronization episodes if
their durations exceed $T$ seconds (here $T=30$ s).}
\label{fig:1}\end{figure}

Our algorithm for the detection of phase synchronization episodes is based on the 
study of cardio-respiratory synchrograms \cite{tutorial,Schaefer,Toledo,Stefanovska}.
For each record, the times $t_k$ of heartbeats are mapped on the continuous cumulative 
phase $\Phi_r(t)$ of the respiratory signal, which we obtain by unfolding the phase 
of the Hilbert transform of the normalized thermistor recording.  Figure 1(a) shows 
a representative synchrogram, where $\Phi_r(t_k) \bmod 2\pi $ is plotted versus $t_k$.  
In case of $n$:1 synchronization (i.~e., if $n$ heartbeats fit to one breathing cycle) 
one observes $n$ parallel horizontal lines in the synchrogram [$n=3$ in Fig. 1(a)].  
In general, to find different ratios $n$:$m$ of phase synchronization, we plot 
$\Psi_m(t_k)=\Phi_r(t_k) \bmod 2\pi m$ versus $t_k$.

While most earlier work relies on a visual evaluation of the synchrograms 
\cite{Schaefer,Stefanovska}, we detect the episodes in a fully 
systematic way.  For each synchronization ratio $n$:$m$ we first replace 
the $n$ phase points $\Psi_m(t_k)$ in each $m$ respiratory cycles by the averages 
$\overline{\Psi}_m(t_k)$ calculated over the corresponding points in the time windows 
from $t_k-\tau/2$ to $t_k+\tau/2$ [Fig. 1(b)]. In the second step, the algorithm 
deletes all phase points $\overline{\Psi}_m (t_k)$ where the mean standard deviation 
of the $n$ points in each $m$ breathing cycles, $\langle \sigma \rangle_n$, is larger 
than $2m\pi/n\delta$.  In the third step, only the phase points $\overline{\Psi}_m(t_k)$ 
in uninterrupted sequences of durations exceeding $T$ seconds are kept [Fig. 1(c)].  

\begin{figure} \begin{center} \epsfig{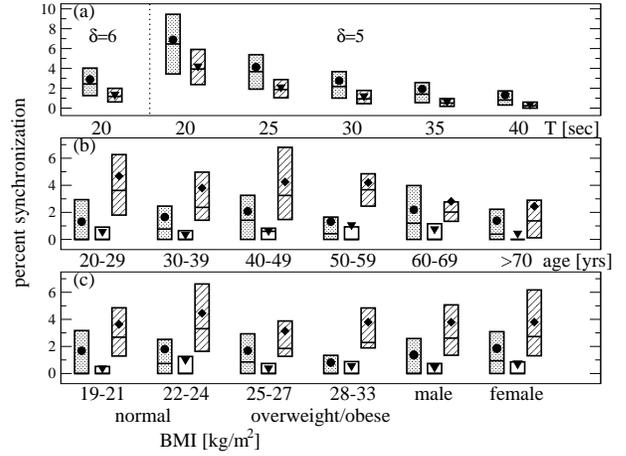}
\end{center} \caption{Medians, upper and lower quartiles (bars) and means (filled symbols)
for the detected synchronization rates (a) versus $T$ for all original data (dotted
bars and circles) and surrogate data (striped bars and triangles), (left of dotted line
$\delta=6$, right of dotted line $\delta=5$); (b,c) versus age,
body mass indices (BMI) and gender, for wakefulness (dotted bars and circles), REM
sleep (blank bars and triangles) and non-REM sleep (striped bars and diamonds)
for $T=30$ s. Note the similar synchronization behavior in all groups.}
\label{fig:3}\end{figure}

Figure 2(a) shows a comparison of the detected synchronization rates in real data
and in surrogate data for several values of the parameter $T$.
The ratio of the mean percentage of synchronization in real data over the mean 
percentage in surrogate data increases from 1.6 for $T=20$ s to 3.4 for $T=40$ s.  
There is no generic limit for $T$.  However, we choose $T=30$ s in order to keep the 
number of arbitrary detection of synchronization at an acceptable low level 
($\approx 1$\% in the surrogate data) while at the same time still detecting
synchronization episodes during all kinds of sleep stages.  In addition, for $T=\tau=
30$ s we have only one effective time scale parameter, which is identical with the
standardized time frame used for the detection of sleep stages \cite{rechtschaffen}.
We note that changing the parameter $\delta$ has a similar effect on the results as 
changing $T$, and we chose $\delta=5$ based on similar considerations as for $T$.
Our results do hardly depend on the duration $\tau$ of the initial running average.

When studying the percentages of synchronized episodes in real data separately
during wakefulness, REM sleep, and non-REM sleep
we obtain a highly significant difference in the frequency of cardio-respiratory
synchronization between the two major physiological states during sleep.
We find 3.8\% synchronization in non-REM sleep compared with just 0.6\% in REM sleep
-- a difference by a factor of 6.3.  Wakefulness is clearly intermediate, since we
find 1.6\% for it.  Similar differences are observed for other values of $T$ and 
$\delta$.  Since our data base contains records of 112 subjects, we can also study 
synchronization separately for several age groups, several groups with
different body mass index (BMI), and men and women.  Figures 2(b,c) show that the
results, for wakefulness, REM sleep, and non-REM sleep are practically the same
for both genders, all BMI groups, and all age groups, although both, heart rate 
and breathing rate are known to depend on BMI and age.  These results prove that 
our finding of significant differences between the cardio-respiratory 
synchronization in REM and non-REM sleep is very stable.

Similar stable differences between REM and non-REM sleep were found in the correlation 
properties of both heartbeat \cite{HBcor} and breathing \cite{BRcor} fluctuations, 
but hardly in the magnitude of these fluctuations \cite{fn1}.  The differences 
were attributed to the influence of the central nervous system with its sleep stage 
regulation in higher brain regions on the autonomous nervous system.  The similarity 
leads us to suggest that the diminished synchronization during REM sleep is also 
caused by influences of the central nervous system.  As long as the heartbeat oscillator 
and the breathing oscillator (as parts of the autonomous nervous system) are 
only affected by uncorrelated noise from higher brain regions, they run like 
two weakly coupled oscillators -- and they clearly show synchronization as expected, 
possibly enhanced by the noise \cite{noise}.  However, if the higher brain regions are 
more active and impose long-term correlated noise on the two oscillators, as is the 
case during REM sleep, the noise disturbs the emergence of synchronized patterns,
leading to a drastic reduction of synchronization episodes.  Hence we suggest from the 
experimental data that correlated noise is rather suppressing synchronization while 
uncorrelated noise might increase it.

Our interpretation is consistent with the result that cardio-respiratory synchronization 
is enhanced in heart transplanted patients, where correlated signals from the brain can 
hardly affect the heartbeat oscillator \cite{Toledo}.  Hence, it supports that any 
relation of the synchronization patterns with cardiac impairments can only be an 
indirect one as reported recently \cite{Hoyer}.  Diminished long-term correlated 
regulation activity might explain the increase of synchronization in well-trained 
athletes \cite{Schaefer}, where fluctuations of heartbeat and breathing might be 
avoided to optimize the cardiovascular system for optimal performance.

\begin{figure} \begin{center} \epsfig{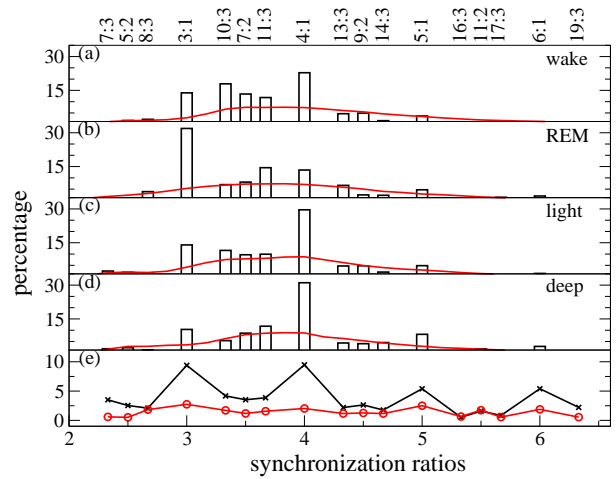}
\end{center} \caption{Normalized histograms of all detected synchronization ratios
as well as distributions of the frequency ratios between heartbeat and breathing
during (a) wakefulness, (b) REM sleep, (c) light sleep, and (d) deep sleep.
Note that 4:1 synchronization is remarkably increased in light sleep and deep sleep,
whereas in REM sleep 3:1 synchronization dominates. In (e) the quotient of the
synchronization ratio histogram with the distribution of the frequency ratios
is shown for the original data (crosses) as well as for the surrogate data (circles).}
\label{fig:3}\end{figure}

In order to gain insight into the mechanism of cardio-respiratory phase synchronization
we have studied, again for all 112 subjects, the distribution of the synchronization
ratios $n$:$m$, where $n$ cardiac cycles are synchronized with $m$ breathing cycles.
Figure 3 shows the normalized histograms of the synchronization ratios during (a)
wakefulness, (b) REM sleep, (c) light sleep, and (d) deep sleep.  The underlying 
continuous curves in Fig. 3(a-d) show the distributions of the frequency ratios between 
heartbeat and breathing independent of synchronization.

To clarify the efficiency of the synchronization mechanism, we show the quotient 
of the synchronization rate histogram with the distribution of the frequency ratios 
(crosses) in Fig. 3(e). When comparing with the corresponding curve for surrogate 
data (circles) it becomes obvious that $n$:1 synchronization is preferred.  This 
involves not just the common ratios of 3:1, 4:1, and 5:1, but also clearly 6:1 
synchronization.  In particular, we do not find any indication of a suppression of 
4:1 synchronization as has been reported recently based on modelling \cite{Kotani}.
This suggests that the feedback from baroreceptors to the respiratory centers
introduced in the model is probably not very important in the healthy subjects
we studied.  The synchronization ratios $n$:2 and $n$:3 are weakly efficient if 
$n$ is low, but not efficient at all for large values of $n$.  Hence, Fig.~3(e) 
proves that cardio-respiratory synchronization is nearly limited to frequency 
ratios $n$:$m$ with very small $m$, but quite independent of the number of 
heartbeats $n$.

>From this behavior we suggest that the physiological synchronization mechanism
is mainly based on an interaction of the respiratory cycle upon heartbeat and 
not vice versa.  This assumption can explain the weak (or absent) efficiency of 
$n$:$m$ synchronization with $m>1$.  If the heartbeat oscillator gets a synchronizing 
kick at a particular phase of each respiratory cycle and $m>1$, only half or 
even less of the kicks coincide with a heartbeat and thus $n$:$m$ synchronization 
cannot be effective.  The assumption is consistent with the result obtained when
studying the direction of synchronization in children and adults \cite{Rosenblum}
and in a recent model \cite{Smely}.  It is also coherent with the result that
synchronization is enhanced under paced respiration \cite{Prokhorov}.

In conclusion, we have studied cardio-respiratory phase synchronization during
different well-defined physiological stages in sleep for a large data base
of healthy subjects.  We observed clearly reduced synchronization during REM 
sleep and enhanced synchronization during non-REM sleep.  The result is stable
for all studied subgroups of subjects; it is neither affected by gender, nor
by age, nor by BMI.  Since REM and non-REM sleep differ mainly in the type of 
activity of higher brain centers, it seems probable that the differences in
cardio-respiratory synchronization are caused by the more and less long-term 
correlated regulation actions of the brain during REM and non-REM, respectively.  
Heart rate and breathing rhythm generators behave like two weakly coupled 
oscillators, where the coupling direction is from breathing to heartbeat.  They 
become synchronized if uncorrelated noise is imposed from the brain while long-term 
correlated noise disturbs the emergence of the synchronized patterns.  Hence, the 
experimental data suggests that correlated noise is suppressing synchronization 
while uncorrelated noise might increase it.

{\em Acknowledgement:}  We thank Shay Moshel, Meir Plotnik, and Diego Rybski for 
discussions.  This work has been supported by the Deutsche Forschungsgemeinschaft 
(grants KA 1676/3 and PE 628/3), by the Minerva Foundation, by the Israel Science 
Foundation, and by the EU project DAPHNet (grant 018474-2).


\begin{thebibliography}{99}
\bibitem{fn1} We note that there is very little ($\approx 14$\%) increase in the 
amplitude of heartbeat or breathing fluctuations during REM sleep when compared with non-REM 
sleep.  The changes in the synchronization behavior can thus not be due to variations in the 
strength of the influences from the brain.  Rather they must be due to the correlation 
structure imposed by these influences.  Long-term correlations are nearly absent in both, 
heartbeat and breathing during non-REM sleep \cite{HBcor,BRcor}.
\end{thebibliography}
\end{document}